\documentclass{aa}  
\usepackage{color}
\usepackage{graphicx}
\usepackage{lscape}
\usepackage{natbib}
\usepackage{natbib,twoopt}
\usepackage[varg]{txfonts}
\usepackage{url}
\usepackage{xspace}
\usepackage{amssymb}
\usepackage{stmaryrd}
\usepackage{siunitx}
\usepackage{textcomp}
\usepackage{placeins}
\usepackage{float}
\usepackage{footmisc}

\bibpunct{(}{)}{;}{a}{}{,}
\usepackage[breaklinks=true, colorlinks=true, linkcolor=blue, citecolor=blue, urlcolor=blue]{hyperref}

\newcounter{Rco}
\newcommand{\Ionst}[1]{\setcounter{Rco}{#1}\Roman{Rco}}
\newcommand{\Ion}[2]{\mbox{#1\,{\scriptsize\Ionst{#2}}}}

\newcommand{\logg}{\mbox{$\log g$}\xspace}

\newcommand{\se}[1]{\mbox{Sect.\,\ref{#1}}}

\newcommand{\Teff}{\mbox{$T_\mathrm{eff}$}\xspace}

\newcommand{\Msol}{$M_\odot$}

\def\Gaia{{\it Gaia}}
\newcommand{\GALEX}{{\it GALEX}}

\begin{document}

\title{Spectroscopic survey of faint planetary-nebula nuclei} \subtitle{VII. Thirty new hydrogen-deficient central stars}

\author{Klaus Werner\inst{1}
    \and Howard E. Bond \inst{2, 3}
    \and Gregory R. Zeimann \inst{4}
    }
    
\offprints{Klaus\,Werner\\ \email{werner@astro.uni-tuebingen.de}}

\institute{
Institut f\"ur Astronomie und Astrophysik, Kepler Center for Astro and Particle Physics, Eberhard Karls Universit\"at, Sand~1, 72076 T\"ubingen, Germany
\and Department of Astronomy and Astrophysics, Penn State University, University Park, PA 16802, USA
\and Space Telescope Science Institute, 3700 San Martin Dr., Baltimore, MD 21218, USA
\and Hobby-Eberly Telescope, University of Texas at Austin, Austin, TX 78712, USA
}

\date{Received 17 March 2026 / Accepted 01 April 2026}

\titlerunning{Hydrogen-deficient planetary-nebula nuclei}

\abstract{Our ongoing spectroscopic survey of faint planetary-nebula nuclei (PNNi) has revealed 30 new hydrogen-deficient central stars. The majority of them (21) belong to the PG1159 spectral class (having He-C-O--dominated atmospheres). They increase the number of known PN central stars of this type from 25 to 46. Our spectral analysis finds that their effective temperatures are high (\Teff = 110\,000--180\,000\,K), locating them in the GW~Vir pulsational instability strip. Future photometric observations should therefore substantially increase the number of known PNN pulsators (currently, it is nine). We found six new members of the O(He) spectral type (with He-dominated atmospheres; \Teff = 120\,000--150\,000\,K), tripling the number of known PNNi of this class. Finally, we identified three hot helium-rich white dwarfs with traces of carbon and/or nitrogen (spectral type DOZ; \Teff = 70\,000--100\,000\,K). They are the first objects of this spectral class found to be associated with a planetary nebula.}


\maketitle
\nolinenumbers

\section{Introduction}
\label{sect:intro}

This is the seventh in a series of papers presenting results from a spectroscopic survey of nuclei of faint Galactic planetary nebulae (PNe). The survey is carried out with the second-generation Low-Resolution Spectrograph (LRS2; \citealt{Chonis2016}) of the 10-m Hobby-Eberly Telescope (HET; \citealt{Ramsey1998,Hill2021}), located at McDonald Observatory in west Texas, USA\null. An overview of the project, a description of the instrumentation and data-reduction procedures, target selection, and some initial results were presented in our first paper \citep[][hereafter Paper~I]{Bond2023a}. There, and in Paper~III \citep{WernerBondIII2024}, we announced the discovery of a total of nine new extremely hot hydrogen-deficient central stars, and we reported on a total of 17 new H-rich nuclei in Paper~VI \citep{Reindl2024}, considerably increasing the number of hot, H-rich objects for which non-local-thermodynamic-equilibrium (NLTE) atmospheric parameters are available. Three other publications in the series have discussed individual objects of special interest. In the present paper, we report the discovery and analysis of 30 additional hot and hydrogen-deficient central stars.

Planetary-nebula nuclei (PNNi) are remnants of low- and intermediate-mass stars that left the asymptotic giant branch (AGB) and are in the process of becoming white dwarfs (WDs). These post-AGB stars contract and heat up, and upon reaching an effective temperature of \Teff $\approx$ 30\,000\,K, their ultraviolet (UV) flux becomes strong enough to ionize the gas envelope, consisting of material shed from the star during its AGB phase -- thus producing a PN\null.  Continued contraction and heating to a maximum effective temperature of more than 100\,000\,K turns the stars into pre-WDs, which eventually enter the WD cooling sequence upon reaching a surface gravity higher than $\logg \approx 7$. Canonical evolutionary theory \citep[e.g.,][]{Kippenhahn2013} predicts that the stellar envelope remains hydrogen-rich throughout this process. Nonetheless, about one third of central stars have been found to be hydrogen-deficient \citep{Weidmann2020}. Several classes of H-deficient PNNi have been established, in order to characterize their diverse stellar spectra. These classes also comprise similar field objects that are not associated with a PN, either because, for example, the nebula has already dispersed, or because the star is not the outcome of single-star evolution.

The PG1159 spectral class (named after their prototype, PG\,1159$-$035 = GW~Virginis) was defined for hot hydrogen-poor (pre-)WDs whose optical spectra are dominated by \Ion{He}{2} and \Ion{C}{4} lines \citep{mcgraw1979,Liebert1980,Bond1984}. Model-atmosphere analyses revealed that their temperatures and gravities range between \Teff = 60\,000--250\,000\,K and \logg = 4.8--8.3 \citep[e.g.,][and subsequent work]{WernerHeberHunger1991}. Their atmospheric abundances are usually dominated by helium and carbon, but often with an admixture of considerable amounts of oxygen. As a typical example, the composition of the prototype is He = 0.33, C = 0.50, and O = 0.17 (mass fractions). The majority of PG1159 stars (in particular all those lying within a PN) are thought to be the result of a late or very late thermal pulse (LTP or VLTP, respectively); see, for example, \cite{werner2006}. These events denote the re-ignition of He-shell burning after the star has left the AGB\null. In the LTP case, the thermal pulse occurs during the transition phase between the AGB and the top of the WD cooling sequence. Helium-shell-flash--driven convection reduces the hydrogen abundance in the envelope by dilution with He- and C-rich interior material. In the VLTP case, helium-shell ignition occurs only during WD cooling, causing ingestion and violent burning of hydrogen. Up until now, 71 PG1159 stars are known, of which 25 are PN central stars.\footnote{From an unpublished list of PG1159 stars maintained by K.W., based on the list published by \cite{werner2006}\label{w06_footnote}.}

In a classification system introduced by \citet{Mendez1986}, the O(He) spectral class consists of pre-WD stars that have helium-dominated atmospheres.\footnote{We note that a spectral class, ``O(C),'' was also introduced by \cite{Mendez1986}, and is occasionally cited in the literature. Since it is identical to the PG1159 class, we retain the latter usage here.} They can occasionally show trace amounts of C, N, and\slash or O (abundances less than about 1\%), but their optical spectra are dominated by \Ion{He}{2} absorption lines with weak or no metal lines. To date, 14 O(He) stars are known \citep[for the latest discovery, see][]{werner2025}. Their  effective temperatures and gravities cover the ranges \Teff = 80\,000--200\,000\,K and \logg = 5.0--6.7, so they coexist in the region of PG1159 stars in the ``Kiel diagram'' (\logg\ versus \Teff). Only three PNNi have until now been assigned to the O(He) spectral class; namely, those of the PNe LoTr~4, K~1-27, and Pa~5 \citep{Reindl2014,DeMarco2015}. The reason for the surface chemistry of these stars is debated. They are often linked to R~Coronae Borealis and extreme helium stars, suggesting that they may result from mergers of helium WDs \citep{Reindl2014}. This scenario, however, cannot explain the formation of PNe around O(He) stars. 

A DO spectral type is assigned to WDs that have spectra dominated by absorption lines of ionized helium \citep[e.g.,][]{Wesemael1993}. Their progenitors are likely the O(He) pre-WDs \citep{Rauch1998} and\slash or PG1159 stars \citep{Liebert1980}. At their lower temperatures, gravitational settling has removed the heavy elements from their atmospheres. 
The hottest DOs retain trace metals in their photospheres by radiative levitation, such that weak metal lines (usually from \Ion{C}{4}) are detectable in their optical spectra. In this case they are classified as DOZ\null. 
The prototype star of the DO class is PG\,1034+001 \citep{Wesemael1993, Wesemael1985}, which has been reported to be the central star of an angularly large and  nearby PN discovered by \citet{Hewett2003}. However, \cite{Frew2013} have argued convincingly that the nebula (Hewett~1) is not a PN, but is instead a Str\"omgren zone in the interstellar medium (ISM) ionized by the hot star \cite[\Teff = 115\,000\,K;][]{wernerrauch2017}. Thus -- until now -- there has been no known DO-type central star of a PN\null. 

In the paper at hand, we report our discoveries and atmospheric analyses of 30 new hot H-deficient PNNi, belonging to spectral types PG1159 (21 objects), O(He) (six objects), and DOZ (three objects). For completeness, we mention that many H-deficient PNNi belong to a fourth spectral class and have Wolf-Rayet--type spectra with emission features; they are denoted as [WR] stars, the brackets distinguishing them from massive Population~I WR stars. The [WR] objects may be progenitors of PG1159 stars \citep{Mendez1986}, but this issue is debated \citep[e.g.,][]{2024RMxAA..60..227H}. For a recent review of the evolution of hydrogen-deficient stars and PNNi, see \cite{MMM2024}.

In \se{sect:targets} we introduce the targets analyzed in this paper, and in \se{sect:observations} we give an overview of our spectroscopic observations. The spectral classifications and analyses are presented in \se{sect:analysis}. 
{We discuss our results in the context of the evolutionary status of the objects in \se{sect:discussion}, and conclude and summarize in \se{sect:summary}.}

\section{Targets}
\label{sect:targets}

Table~\ref{tab:targetlist} lists the 30 PNNi that our survey observations revealed to be hot, hydrogen-deficient stars. 
The first two columns give the names and PN\,G designations of the host PNe; these are taken from the online Hong-Kong/AAO/Strasbourg/H$\alpha$ Planetary Nebulae (HASH) database\footnote{\url{http://hashpn.space/}} \citep{Parker2016, Bojicic2017}. The next six columns list the celestial coordinates, parallaxes, and magnitudes and colors of the central stars, as given in \Gaia\/ Data Release~3 (DR3; \citealt{Gaia2016, Gaia2023}).\footnote{\url{https://vizier.cds.unistra.fr/viz-bin/VizieR-3?-source=I/355/gaiadr3}} The final column in Table~\ref{tab:targetlist} lists the angular radii of the PNe, taken from HASH\null. Further information about the objects, including direct images of the PNe at several wavelengths, is also available from HASH\null. As can be inferred from their names, many of these PNe were discovered in recent years by amateur astronomers, mostly through an examination of publicly available sky surveys, and they generally have very low surface brightnesses. Several of these discoveries have been followed up by amateurs who obtain very deep and often remarkable images, as is mentioned below.

\begin{table*}
\centering
\caption{PN nucleus target list, \Gaia\/ DR3 data, and angular radii.}
\label{tab:targetlist}
\begin{tabular}{lccccccccc}
\hline\hline
\noalign{\smallskip}
Name 
& PN G
& RA (J2000)
& Dec (J2000)
& Parallax [mas]
& {$G$}
& $G_{\rm BP}-G_{\rm RP}$
& $R_{\mathrm{PN}}$ \\
\noalign{\smallskip}
\hline
\noalign{\smallskip}
Abell 52    & 050.4+05.2   & 19 04 32.335 & +17 57 07.69   & $0.338 \pm 0.114$ & 17.68 & +0.15   & $\phantom{0}18.5''$ \\ 
Alves 5     & 022.9+22.7   & 17 12 41.636 &     +01 55 33.22   & $1.130 \pm 0.124$ & 17.29 & $-$0.14 & $180''$   \\
Dr 37       & 046.9$-$00.8 & 19 20 26.581 & +12 02 19.59   & $0.443 \pm 0.180$ & 18.79 & +1.41   & $60''$    \\
Ek 3        & 076.8$-$03.7 & 20 41 27.636 & +35 53 52.88   & $0.892 \pm 0.142$ & 18.53 & $-$0.04 & $120''$   \\
Fal 3       & 057.0$-$03.7 & 19 51 18.070 & +19 25 09.32   & $0.603 \pm 0.075$ & 16.97 & $-$0.09 & $141''$   \\
Fal Objet 1 & 219.6$-$01.3 & 06 58 50.283 & $-$06 33 32.96 & $0.666 \pm 0.051$ & 15.96 & $-$0.04 & $1410''$  \\
HaWe 15     & 099.7$-$08.8 & 22 30 33.433 &     +47 31 23.29   & $0.830 \pm 0.122$ & 18.25 & $-$0.34 & $147.5''$ \\
K 1-17      & 051.5+06.1   & 19 03 37.329 & +19 21 22.79   &        (1)        & 18.29 & +1.03   & $56''$    \\
K 2-1       & 173.7$-$05.8 & 05 07 08.307 & +30 49 18.51   & $0.448 \pm 0.172$ & 18.09 & +0.04   & $63''$    \\
Kn 58       & 136.8$-$13.2 & 02 12 27.839 & +47 27 10.13   & $0.152 \pm 0.344$ & 19.45 & $-$0.36 & $37.5''$  \\
Kn 62       & 175.6+11.4   & 06 23 55.423 & +38 15 14.50   & $0.252 \pm 0.191$ & 18.55 & $-$0.16 & $63''$    \\
Kn 63       & 200.5$-$13.1 & 05 42 06.707 & +04 43 02.67   & $0.767 \pm 0.095$ & 17.47 & $-$0.01 & $176''$   \\
Kn 121      & 058.8$-$16.9 & 20 42 01.940 & +13 51 14.91   & $0.783 \pm 0.043$ & 15.57 & $-$0.46 & $180''$      \\
NGC 6765    & 062.4+09.5   & 19 11 06.558 & +30 32 43.67   & $0.275 \pm 0.078$ & 17.60 & $-$0.17 & $20''$    \\
NHZ 2       & 046.6$-$16.5 & 20 15 08.522 & +04 08 58.28   & $0.507 \pm 0.129$ & 17.72 & $-$0.46 & $250''$   \\
Ou 7        & 068.3+02.6   & 19 52 45.354 & +32 25 40.25   & $0.551 \pm 0.110$ & 18.23 & +0.10   & $171''$   \\
Pa 28       & 076.8$-$08.1 & 20 58 10.952 & +33 08 33.24   & $0.295 \pm 0.163$ & 18.83 & $-$0.35 & $116.5''$ \\
Pa 144      & 050.2$-$11.9 & 20 06 31.812 & +09 26 20.84   & $0.087 \pm 0.182$ & 18.42 & $-$0.38 & $13.5''$     \\
Pa 146      & 058.5$-$13.3 & 20 29 09.414 & +15 37 00.63   & $0.327 \pm 0.088$ & 17.16 & $-$0.36 & $56.5''$     \\
Pa 153      & 155.9+07.7   & 05 09 07.623 & +53 10 28.18   & $0.436 \pm 0.122$ & 17.93 & +0.38   & $83.5''$  \\
Pa 180      & 093.1+05.3   & 21 00 00.225 & +54 18 11.06   & $2.009 \pm 0.069$ & 17.49 & $-$0.27 & $30''$    \\
Pa J0637+3327&181.2+11.8   & 06 37 28.264 & +33 27 07.31   & $0.952 \pm 0.111$ & 17.16 & $-$0.41 & $>480''$  \\
PFP 1       & 222.1+03.9   & 07 22 17.697 & $-$06 21 45.98 & $1.872 \pm 0.074$ & 15.82 & $-$0.57 & $575''$   \\ 
StDr 13     & 204.4$-$00.4 & 06 34 22.654 & +07 22 20.12   & $0.837 \pm 0.162$ & 18.08 & $-$0.07 & $225''$   \\
StDr 29     & 065.4+20.2   & 18 29 02.076 & +37 20 05.66   & $1.371 \pm 0.088$ & 17.80 & $-$0.50 & $150''$   \\
StDr 61     & 134.3+02.5   & 02 35 23.865 & +63 03 02.17   & $3.994 \pm 0.037$ & 15.55 & $-$0.33 & $390''$   \\
StDr 144    & 179.2$-$01.6 & 05 37 24.764 & +28 42 39.42   & $1.315 \pm 0.083$ & 16.94 & +0.18   & $480''$   \\
StDr 162    & 162.7+03.2   & 05 11 01.983 & +45 02 24.54   & $0.425 \pm 0.194$ & 18.76 & +0.06   & $171''$   \\
TaWe 1      & 208.9$-$07.8 & 06 16 15.384 & $-$00 00 25.03 & $0.592 \pm 0.194$ & 18.57 & $-$0.21 & $72.5''$  \\ 
WHTZ 1      & 040.8$-$09.7 & 19 40 43.840 & +02 30 31.93   & $0.555 \pm 0.071$ & 16.89 & $-$0.10 & $91.5''$  \\ 
\noalign{\smallskip}
\hline
\end{tabular} 
\tablefoot{Target names and angular radii, $R_{\mathrm{PN}}$, of PNe taken from HASH database. 
\tablefoottext{1}{No parallax available in \Gaia\/ DR3. 
} }

\end{table*}

In the following paragraphs, we give brief details of the discoveries of these faint PNe and their nuclei. In most cases, the central star is obvious through an inspection of sky surveys, especially color images from the Pan-STARRS survey.\footnote{\url{https://ps1images.stsci.edu/cgi-bin/ps1cutouts}}  However, because of the high effective temperatures of the central stars, many of them also appear in catalogs of WDs and hot subdwarfs, but often without being recognized as PNNi. Most of the stars have been detected in the UV by the {\it Galaxy Evolution Explorer\/} (\GALEX\/) survey \citep[see][henceforth ``B+17'']{Bianchi2017}, in those cases in which the sites were imaged. To our knowledge, all but one of our targets have not previously had their spectra discussed in the literature; for example, they are not contained in the compilation of spectral classifications of central stars assembled by \citet{Weidmann2020}. The one exception is the nucleus of NGC~6765, discussed below.

\begingroup

\setlength{\parskip}{-12pt}

\paragraph{Abell 52} The PN and its central star were identified in the classical study of the Palomar Observatory Sky Survey (POSS) by \cite{Abell1966}. The star is a \GALEX\/ source with magnitudes FUV = 17.2 and NUV = 17.7 (B+17). It appeared in the list of hot 
\hbox{subdwarfs} of \cite{Geier2019}, henceforth ``G+19,'' and in the WD catalog by \cite{GentileFusillo2019}, henceforth ``GF+19.'' It was also identified in the \Gaia\/ source catalog of PNNi by \cite{Chornay2020}, henceforth ``CW20.''
\footnote{A recent image by US amateur Jerry Macon is here: \url{https://app.astrobin.com/i/xvq9h1}.}

\paragraph{Alves 5} The discovery of this nebula by Portuguese amateur Filipe Alves was announced by \cite{LeDu2022}, henceforth ``LD+22,'' and classified as a likely PN\null. The central star is a \GALEX\/ source with magnitudes FUV = 15.7 and NUV = 16.6 (B+17). It appeared in the WD catalog by GF+19.

\paragraph{Dr 37} The discovery of this PN candidate by German amateur Marcel Drechsler was announced in LD+22. We identified a candidate PNN, \Gaia\/ source number 4315941098777488768, through an inspection of Pan-STARRS images. This star stands out as being bluer than the surrounding objects; however, there is heavy interstellar extinction at the site, which explains its red color in the \Gaia\/ system of $G_{\rm BP}-G_{\rm RP}=+1.41$.


\paragraph{Ek~3} The discovery of this PN candidate by Swedish amateur Sven Eklund is noted on a PN website maintained by the French amateurs Pascal Le D\^u and Thomas Petit (planetarynebulae.net\footnote{\url{https://planetarynebulae.net/EN/index.php}}). The central star appeared in the list of hot subdwarfs of G+19 and in the WD catalog by GF+19.

\paragraph{Fal~3} This is a PN candidate discovered by US amateur Bray Falls (see planetarynebulae.net), and kindly pointed out to us by Sakib Rasool.\footnote{Falls' discovery imaging is available at \url{https://app.astrobin.com/i/v4znmz}.} We identified a candidate blue PNN, which also appeared in the list of hot subdwarfs of C+22 and in the WD catalog by GF+19.

\paragraph{Fal Objet 1} This is another new candidate PN discovered by Falls (planetarynebulae.net).\footnote{The discovery image is here: \url{https://app.astrobin.com/i/wr5dh1}.} This object should not be confused with a different PN named ''Fal~1.'' We identified a candidate central star, which is also listed in the hot subdwarf catalog of G+19. 

\paragraph{HaWe~15} The discovery of this PN was first announced as the 13th object in \cite{HDW13}, and is therefore also known in the literature under the name HDW~13. The central star was identified in the \Gaia\/ source catalogs by CW20 and is a \GALEX\/ source with magnitude NUV = 17.2 \citep[][henceforth ``G+23'']{Gomez2023}.\footnote{A recent image by US amateur Kevin Quin is here: \url{https://ssr.app.astrobin.com/i/8ar8ek}.}

\paragraph{K 1-17} The PN was discovered by \cite{Kohoutek1963}. 
{ The central star ($G = 18.29$) was identified in the \Gaia\/ source catalogs by \cite{Gonzalez2021}. There is a fainter ($G = 20.46$) nearby ($0\farcs6$) star, which in the literature is sometimes misidentified as the central star (e.g., in CW20). Our spectrum shows signatures of a cool companion (see comment in Sect.\,\ref{sect:observations}). Whether they stem from the nearby star, or from an unresolved companion, remains unclear.} The spectral-energy distribution exhibits an IR excess. The Pan-STARRS $griz$ magnitudes are 18.3, 17.9, 17.7, and 17.5, respectively, and the 2MASS $JHK$ magnitudes are 16.3, 15.8, and 15.6, respectively.\footnote{\url{https://vizier.cds.unistra.fr}} The central star is a \GALEX\/ source with magnitudes FUV = 18.9 and NUV = 19.0 (B+17).\footnote{A recent image of the PN by the British amateur Peter Goodhew is here: \url{https://www.imagingdeepspace.com/kohoutek-1-17.html}.}

\paragraph{K 2-1} This PN was also identified by \cite{Kohoutek1963}, but appeared first as a reflection nebula in a list of diffuse Galactic nebulae by \cite{Struve1962}. The central star is a \GALEX\/ source with magnitudes FUV = 16.6 and NUV = 16.9 (B+17). It is listed in the WD catalog of \cite{GentileFusillo2021}.\footnote{A recent image by Goodhew is here: \url{https://www.imagingdeepspace.com/k2-1.html}.}

\paragraph{Kn 58} This nebula was discovered by the Austrian amateur Matthias Kronberger and coworkers, and announced as a new PN candidate in \cite{Kronberger2012}. It was confirmed as a true PN by \cite{Ritter2023}. The central star is listed as a hot subdwarf in \cite{Culpan2022}, henceforth ``C+22.'' It is a \GALEX\/ source with magnitude NUV = 18.3 (G+23).

\paragraph{Kn 62} This object was announced as a new PN candidate by \cite{Kronberger2014} and confirmed as a true PN by \cite{Ritter2023}. The central star is in the list of hot subdwarfs of G+19 and in the WD catalog by GF+19. It is a \GALEX\/ source with magnitudes FUV = 17.36 and NUV = 17.8 (B+17).

\paragraph{Kn 63} This PN candidate was also announced by \cite{Kronberger2014}. A deep image by Goodhew confirms its PN nature.\footnote{\url{https://www.imagingdeepspace.com/kn-63.html}} The central star is in the list of hot subdwarfs of G+19 and in the WD catalog by GF+19.

\paragraph{Kn 121} The discovery of this PN was announced by LD+22, and it was classified as a true PN\null.\footnote{An image by Goodhew is here: \url{https://www.imagingdeepspace.com/kn-121.html}.} The central star is in the list of hot subdwarfs of G+19 and in the WD catalog by GF+19. It is a \GALEX\/ source with magnitudes FUV = 14.1 and NUV = 14.4 (B+17). 

\paragraph{NGC 6765} The central star of this classical PN appeared in the WD catalog of \cite{McCook1999} and in the hot subdwarf catalog by C+22. It was identified in the \Gaia\/ source catalogs by CW20.\footnote{A recent image of the PN by  Goodhew can be found here: \url{https://app.astrobin.com/u/PeterGoodhew?i=0ker9f}.} Based on a relatively poor spectrum, the central star was tentatively classified as PG1159 by \cite{Napiwotzki1995}, which we confirm with our new observations presented here. 

\paragraph{NHZ 2} This PN candidate announced by the ``New Horizon Team'' of amateur astronomers is listed on planetarynebulae.net. The central star is in the list of hot subdwarfs of G+19 and in the WD catalog by GF+19. It is a \GALEX\/ source with magnitudes FUV = 15.9 and NUV = 16.6 (B+17).

\paragraph{Ou 7} The discovery of this nebula was announced by \cite{LeDu2017}. The nebula was classified as a true PN by LD+22.\footnote{Here is the discovery image by Nicolas Outters: \url{https://www.outters.fr/?p=3876}.} The central star appeared in the list of hot subdwarfs of C+22 and in the WD catalog by GF+19. 

\paragraph{Pa 28} This object, discovered by US amateur Dana Patchick, was announced as a new PN candidate by \cite{Kronberger2014}. It was confirmed as a true PN by \cite{Ritter2023}.\footnote{An image by Goodhew is here: \url{https://www.imagingdeepspace.com/pa28.html}.} The central star is in the list of hot subdwarfs of G+19 and in the WD catalog by GF+19. It is a \GALEX\/ source with magnitudes FUV = 17.9 and NUV = 18.0 (B+17).

\paragraph{Pa 144} The discovery of this object was announced by LD+22 and it was classified as a true PN. The central star is in the list of hot subdwarfs of G+19 and in the WD catalog by GF+19. It is a \GALEX\/ source with magnitude NUV = 17.5 (G+23).

\paragraph{Pa 146} This is a possible PN listed in the HASH database, discovered by Patchick. The central star appeared in the list of hot subdwarfs of G+19, in the WD catalog of \cite{GentileFusillo2021}, and is a \GALEX\/ source with magnitudes FUV = 15.5 and NUV = 16.1 (B+17).

\paragraph{Pa 153} The object is listed in the HASH database and classified as a likely PN\null.\footnote{An image by Goodhew is here: \url{https://www.imagingdeepspace.com/pa-153.html}.} The candidate central star at coordinates given by HASH is not listed in SIMBAD\null. Its \Gaia\/ name is DR3 266157955103568256. It is a \GALEX\/ source with magnitudes FUV = 19.5 and NUV = 19.6 (G+23). Its red color ($G_{\rm BP}-G_{\rm RP}=+0.38$) is due to significant interstellar extinction. 

\paragraph{Pa 180} The discovery of this PN candidate was announced by LD+22. The central star appears in the WD candidate catalog by \cite{GentileFusillo2015}. Based on \Gaia\/ XP spectra, \cite{Vincent2024} tentatively classified the star as a DA WD with \Teff = 29\,848\,K and \logg = 7.27. However, as we show below, it is considerably hotter.

\paragraph{Pa J0637+3327} The PN was discovered by Patchick (see HASH database). We thank Rasool for encouraging us to include it in our target list.\footnote{An image from a group of amateurs led by Jon Talbot is here: \url{https://app.astrobin.com/i/syhc4r?r=0}.} The central star is in the list of hot subdwarfs of G+19 and in the WD catalog by GF+19. It is a \GALEX\/ source with magnitudes FUV = 15.5 and NUV = 16.1 (B+17).

\paragraph{PFP~1} This PN was discovered and studied in detail by \cite{PFP2004}.\footnote{An image by Italian amateur Marco Lorenzi is here: \url{https://www.glitteringlights.com/search\#q=pfp+1}, and another one by German amateur Andreas Bringmann here: \url{https://app.astrobin.com/i/394663}.} 
The central star is in the list of hot subdwarfs of G+19, and in the WD catalog by GF+19. It was identified in the \Gaia\/ source catalogs by CW20. Based on \Gaia\/ XP spectra, \cite{Vincent2024} tentatively classified the star as a DO WD with \Teff = 122\,373\,K and \logg = 7.612.

\paragraph{StDr 13} The discovery of this PN by the French-German amateur team Xavier Strottner and Marcel Drechsler was announced by LD+22. The object was classified as a true PN.\footnote{An image by  Goodhew is here: \url{https://www.imagingdeepspace.com/stdr-13-with-stdr-155.html}.} The central star is in the list of hot subdwarfs of G+19, and in the WD catalog by GF+19. It is a \GALEX\/ source with magnitude NUV = 18.4 (G+23).

\paragraph{StDr 29} The discovery of this nebula was announced by LD+22. The object was classified as a likely PN\null.\footnote{Here is an image by Goodhew: \url{https://www.imagingdeepspace.com/stdr-29.html}.}  The central star is in the WD catalog by GF+19. It is a \GALEX\/ source with magnitudes FUV = 15.9 and NUV = 16.3 (B+17).

\paragraph{StDr 61} The discovery of this nebula was announced by LD+22 and classified as a possible PN\null.\footnote{An image by French amateur Mathieu Guinot is here: \url{https://guinotmathieu.wixsite.com/astrophotographies/st-dr-61?lang=en}.} The central star is in the WD catalog by GF+19. It is a \GALEX\/ source with magnitudes FUV = 15.9 and NUV = 16.3 (B+17). Based on \Gaia\/ XP spectra, \cite{Vincent2024} classified the star as a DO WD with \Teff = 29\,436\,K and \logg = 6.906. However, we find it to be substantially hotter.

\paragraph{StDr 144} This is a new PN candidate in planetarynebulae.net. It lies superposed on the supernova remnant Simeis~147, the ``Spaghetti” nebula. Near the center of the PN is a blue star that was identified as a UV-bright source, Lanning~658, by \citet{Lanning2004}. This star is also listed in the WD catalog by GF+19. 

\paragraph{StDr 162} is listed as a possible PN on planetarynebulae.net. The central star is in the WD catalog of GF+19. It is a \GALEX\/ source with magnitude NUV = 19.4 (G+23).

\paragraph{TaWe~1} This PN was discovered by \cite{tawe1995} in an investigation of POSS photographic prints.\footnote{A modern deep image of the nebula is available at \url{https://app.astrobin.com/u/Marcel_Drechsler?i=zyam76}.} Its central star is in the list of hot subdwarfs of G+19, and in the WD catalog by GF+19. It was identified in the \Gaia\/ source catalogs by CW20. We again thank Rasool for pointing out this PN.

\paragraph{WHTZ 1} The PN was discovered by \cite{Weinberger1999}, again through searches of the POSS, and studied in detail by \cite{Parker2022}.\footnote{An image by Goodhew can be found here: \url{https://www.imagingdeepspace.com/whtz-14-593513.html}.} The central star appeared in the list of hot subdwarfs of C+22 and in the WD catalog by GF+19. It was identified in the \Gaia\/ source catalogs by CW20.

\endgroup

\section{Spectroscopic observations}
\label{sect:observations}

Paper~I gives full details of the LRS2 instrumentation used for our survey. We note here that LRS2 is composed of two integral-field-unit (IFU) spectrographs: blue (LRS2-B) and red (LRS2-R). All of the observations discussed in this paper were made with LRS2-B, which employs a dichroic beamsplitter to send light simultaneously into two units: the ``UV'' channel (covering 3640--4645~\AA\ at a resolving power of 1910), and the ``Orange'' channel (covering 4635--6950~\AA\ at a resolving power of 1140).

An observation log for our LRS2-B exposures is presented in Table~\ref{tab:observations}.
The data were initially processed using \texttt{Panacea} \citep{Zeimann2026}, which performs bias and flat-field correction, fiber extraction, and wavelength calibration. An absolute-flux calibration was derived from default response curves, measurements of the telescope mirror illumination, and estimates of exposure throughput from guider images. We note that the UV and Orange channels overlap between $\sim$4600--4700~\AA, where the instrumental throughput of both channels exhibits a dip. Small shifts in the effective pivot wavelength between exposures can therefore introduce additional variability in the flux calibration within this overlap region compared to the remainder of the spectrum.

We then applied \texttt{LRS2Multi}\footnote{\url{https://github.com/grzeimann/LRS2Multi}} to the un-sky-subtracted, flux-calibrated fiber spectra to perform background and sky subtraction in an annular aperture, and source extraction using a 2$\arcsec$ radius aperture. When applicable, multiple exposures were combined using inverse-variance weighting. The final spectra from both channels were resampled to a common linear grid with 0.7~\AA\ spacing, and then normalized to a flat continuum for atmospheric analysis. 

\section{Classification and spectral analysis}
\label{sect:analysis}

\subsection{Spectral classification}

We began the analysis of our 30 targets by inferring spectral types from examination of our LRS2-B spectra, based on the classification criteria described in the introduction (see also \citealt{Werner1992}). In
Table~\ref{tab:parameters} the names of the host PNe and the spectral classifications of their nuclei are listed in the first two columns. The next two columns give their atmospheric parameters (effective temperature and surface gravity), and the following four columns the abundances (mass fractions) of He, C, N, and O, all of which were determined as described below. The final column gives stellar masses, also derived as discussed below. The majority (21) of our targets are classified as hot PG1159 stars. According to the classification scheme introduced by \cite{Werner1992}, they belong to the subtypes ``E'' and ``lgE'' because they show \Ion{He}{2} and \Ion{C}{4} emission-line cores within the absorption trough at 4600--4700\,\AA, and some of them additionally indicate a relatively low surface gravity, respectively. The PNN of Ek~3 is the only PG1159 star in our list of subtype ``A,'' i.e., it shows only absorption at the trough, with no emission-line cores. Six of our PNNi are classified as O(He) stars, and three as hot DOZ WDs. 

\begingroup
\begin{table*}
\centering
\caption{Spectral types, atmospheric parameters, chemical abundances (1), and Kiel masses (2) of our program stars.}
\label{tab:parameters}
\begin{tabular}{llccccccc}
\hline 
\hline 
\noalign{\smallskip}
Name & Spectral Type      & \Teff\ [kK]& \logg        & He   & C    & N    & O    & $M$ [$\mathrm{M}_{\odot}$]  \\
\hline
\noalign{\smallskip}
Abell 52     & O(He)      & $130\pm20$ & $5.9\pm0.3$ & 0.99 &  -   & 0.01 &  -   & 0.57\ {\tiny (+0.23,$-$0.06)}  \\           
Alves 5      & PG1159/E   & $150\pm20$ & $7.0\pm0.5$ & 0.47 & 0.48 &  -   & 0.05 & 0.55\ {\tiny (+0.09,$-$0.03)}  \\
Dr 37        & PG1159/lgE & $150\pm20$ & $6.5\pm0.5$ & 0.49 & 0.49 &  -   & 0.02 & 0.55\ {\tiny (+0.16,$-$0.03)}  \\
Ek 3         & PG1159/A   & $110\pm15$ & $7.0\pm0.5$ & 0.78 & 0.22 &  -   &$\leq0.06$&0.51\ {\tiny (+0.08,$-$0.04)}\\
Fal 3        & PG1159/lgE & $150\pm20$ & $6.8\pm0.5$ & 0.47 & 0.48 &  -   & 0.05 & 0.54\ {\tiny (+0.08,$-$0.02)}  \\
Fal Objet 1  & O(He)      & $140\pm20$ & $6.0\pm0.3$ & 0.99 &  -   & 0.01 &  -   & 0.58\ {\tiny (+0.23,$-$0.06)}  \\
HaWe 15      & PG1159/E   & $130\pm20$ & $7.5\pm0.5$ & 0.74 & 0.20 &  -   & 0.06 & 0.59\ {\tiny (+0.19,$-$0.08)}  \\
K 1-17       & PG1159/lgE & $160\pm20$ & $7.0\pm0.5$ & 0.49 & 0.49 &  -   & 0.02 & 0.56\ {\tiny (+0.05,$-$0.03)}  \\
K 2-1        & PG1159/lgE & $150\pm20$ & $6.5\pm0.5$ & 0.47 & 0.48 &  -   & 0.05 & 0.55\ {\tiny (+0.16,$-$0.03)}  \\
Kn 58        & PG1159/E   & $150\pm20$ & $7.0\pm0.5$ & 0.47 & 0.48 &  -   & 0.05 & 0.55\ {\tiny (+0.09,$-$0.03)}  \\
Kn 62        & PG1159/lgE & $150\pm20$ & $6.5\pm0.5$ & 0.47 & 0.48 &  -   & 0.05 & 0.55\ {\tiny (+0.16,$-$0.03)}  \\
Kn 63        & PG1159/E   & $150\pm20$ & $7.0\pm0.5$ & 0.47 & 0.48 &  -   & 0.05 & 0.55\ {\tiny (+0.09,$-$0.03)}  \\
Kn 121       & O(He)      & $150\pm20$ & $6.2\pm0.3$ & 0.99 &  -   & 0.01 &  -   & 0.58\ {\tiny (+0.18,$-$0.06)}  \\
NGC 6765     & PG1159/lgE & $180\pm20$ & $6.5\pm0.5$ & 0.47 & 0.48 &  -   & 0.05 & 0.61\ {\tiny (+0.27,$-$0.05)}  \\
NHZ 2        & PG1159/E   & $130\pm20$ & $7.0\pm0.5$ & 0.47 & 0.48 &  -   & 0.05 & 0.52\ {\tiny (+0.09,$-$0.01)}  \\
Ou 7         & PG1159/E   & $150\pm20$ & $7.0\pm0.5$ & 0.47 & 0.48 &  -   & 0.05 & 0.55\ {\tiny (+0.09,$-$0.03)}  \\
Pa 28        & PG1159/E   & $150\pm20$ & $7.0\pm0.5$ & 0.47 & 0.47 & 0.01 & 0.05 & 0.55\ {\tiny (+0.09,$-$0.03)}  \\
Pa 144       & O(He)      & $130\pm20$ & $5.7\pm0.3$ & 0.99 &  -   & 0.01 &  -   & 0.62\ {\tiny (+0.30,$-$0.10)}  \\
Pa 146       & O(He)      & $140\pm20$ & $6.0\pm0.3$ & 0.99 &  -   & 0.01 &  -   & 0.58\ {\tiny (+0.23,$-$0.06)}  \\
Pa 153       & PG1159/E   & $150\pm20$ & $7.0\pm0.5$ & 0.47 & 0.48 &  -   & 0.05 & 0.55\ {\tiny (+0.09,$-$0.03)}  \\
Pa 180       & DOZ        & $ 70\pm 5$ & $7.5\pm0.3$ & 0.99 & 0.01 &  -   &  -   & 0.51\ {\tiny (+0.07,$-$0.03)}  \\
Pa J0637+3327& PG1159/E   & $130\pm20$ & $7.5\pm0.5$ & 0.74 & 0.20 &  -   & 0.06 & 0.59\ {\tiny (+0.19,$-$0.08)}  \\
PFP 1        & PG1159/E   & $140\pm20$ & $7.5\pm0.5$ & 0.74 & 0.20 &  -   & 0.06 & 0.60\ {\tiny (+0.19,$-$0.08)}  \\
StDr 13      & PG1159/E   & $140\pm20$ & $7.5\pm0.5$ & 0.74 & 0.20 &  -   & 0.06 & 0.60\ {\tiny (+0.19,$-$0.08)}  \\
StDr 29      & DOZ        & $100\pm15$ & $7.7\pm0.3$ & 0.98 &$\leq0.01$&0.02&  - & 0.61\ {\tiny (+0.09,$-$0.06)}  \\
StDr 61      & DOZ        & $ 75\pm 5$ & $7.5\pm0.3$ & 0.97 & 0.03 &  -   &  -   & 0.52\ {\tiny (+0.07,$-$0.03)}  \\
StDr 144     & PG1159/E   & $130\pm20$ & $7.0\pm0.5$ & 0.74 & 0.20 &  -   & 0.06 & 0.52\ {\tiny (+0.09,$-$0.01)}  \\
StDr 162     & O(He)      & $120\pm20$ & $6.0\pm0.3$ & 1.00 &  -   &$\leq0.01$&- & 0.53\ {\tiny (+0.14,$-$0.03)}  \\
TaWe 1       & PG1159/E   & $140\pm20$ & $7.0\pm0.5$ & 0.47 & 0.48 &  -   & 0.05 & 0.54\ {\tiny (+0.09,$-$0.02)}  \\
WHTZ 1       & PG1159/lgE & $180\pm20$ & $6.5\pm0.5$ & 0.47 & 0.48 &  -   & 0.05 & 0.61\ {\tiny (+0.27,$-$0.05)}  \\
\noalign{\smallskip}
\hline
\end{tabular} 
\tablefoot{  
\tablefoottext{1}{Element abundances of best-fit models in mass fractions; see text for their estimated errors.} 
\tablefoottext{2}{Masses determined from the evolutionary tracks displayed in Fig.\,\ref{fig:gteff}.} 
}
\end{table*}
\endgroup

Figures\,\ref{fig:pg1159} and~\ref{fig:ohe} present plots of our rectified LRS2-B spectra. The stars are grouped by spectral classes, and within each class are ordered by decreasing effective temperature. Overplotted are our best-fit models, described in the next subsection, with their parameters and abundances  indicated in the labels above each spectrum. 

\subsection{Model atmospheres}

For the quantitative spectral analysis within the spectral groups, we proceeded as follows: (1)~For the PG1159 stars, we computed a small grid of NLTE model atmospheres of the type introduced by \citet{werner14}. They were calculated using the T\"ubingen Model-Atmosphere Package (TMAP) for plane-parallel models in radiative and hydrostatic equilibrium \citep{wernertmap2003}. The constituents of the models are He, C, and O\null. The grid covers the range \Teff = 100\,000--200\,000\,K in steps of 10\,000\,K, and \logg = 6.0--8.0 in steps down to 0.2\,dex. The abundances (mass fractions) of the chemical elements of the models range between He = 0.47--0.74, C = 0.20--0.49, and O = 0.02--0.06 in different step sizes down to 0.01. One of our PG1159 stars, Pa~28, exhibits \Ion{N}{5} emission lines, so we computed a few models including N as a trace element in subsequent line-formation iterations, meaning the atmospheric structure was kept fixed. (2)~For the O(He) and DOZ stars, we computed smaller grids of pure He models and He+C models, respectively, introducing N as a trace element in a manner similar to the analysis of our PG1159 stars.

\subsection{PG1159 stars}

The spectra of our PG1159 stars display the defining features of \Ion{He}{2}, \Ion{C}{4}, and \Ion{O}{6}, as marked in Figs.\,\ref{fig:pg1159} and \ref{fig:ohe}. Coarse estimates of the stellar effective temperatures are possible just by visual inspection of a few emission and absorption lines. Generally, with increasing effective temperature the \Ion{C}{4}~5801/5812\,\AA\ doublet changes from absorption into emission at around 120\,000\,K \citep[in detail depending also on \logg;][]{WernerHeberHunger1991}. At this temperature, the doublet is therefore barely detectable, or absent. The fact that our PG1159 stars show the feature in emission or lack it means that they must be significantly hotter than 100\,000\,K.

The hottest three stars in our sample are the nuclei of \hbox{K 1-17}, NGC~6765, and WHTZ\,1, as indicated by the presence of the \Ion{O}{6} 3811/3834\,\AA\ doublet in emission (from the photosphere), along with an emission line at 6069\,\AA\null. The latter feature was identified as being due to \Ion{Ne}{8}, in a sample of PG1159 stars by \citet{Werner2007} and in Paper~I; it is seen in objects with temperatures of about 170\,000\,K\null. Another \Ion{Ne}{8} emission line, at 4341\,\AA, is visible in WHTZ~1. However, in NGC~6765, this feature is probably dominated by H$\gamma$ emission from the PN\null. With decreasing temperature, the emission strength of the \ion{O}{vi} doublet becomes weaker, and at about 150\,000\,K, the doublet starts to appear in absorption. Such absorptions are seen in many of our PG1159 stars (e.g., Alves~5 and NHZ~2), which are consequently the coolest PG1159 stars in our sample; they have temperatures of around 130\,000\,K\null. The fact that we cannot clearly identify this doublet in some of our PG1159 stars indicates that they have intermediate temperatures of about 150\,000\,K.

A few of the spectra show anomalous features: (1)~The spectrum of K 1-17 is contaminated by a cool companion star. We identify the \Ion{Mg}{1} $b$ triplet in absorption at 5167--5184\,\AA\ (see also the notes in Sect.\,\ref{sect:targets}). (2)~The spectra of K 2-1 and NGC~6765 show contamination by inadequately subtracted PN emission lines. (3)~In Kn~58, on the other hand, the PN lines are oversubtracted, producing spurious sharp absorption features.

The finally adopted effective temperatures and gravities of the PG1159 stars were determined using the \Ion{He}{2} line profiles, together with the appearance of the \Ion{C}{4} 5801/5812 doublet and the \Ion{O}{6} doublet at 3811/3834\,\AA, and the \Ion{O}{6} feature at 5291\,\AA\null. We consider our error estimates (20\,000\,K and 0.5\,dex for temperature and gravity, respectively) to be rather conservative.

The main indicator for the He/C abundance ratio is the relative strength of the \Ion{He}{2} and \Ion{C}{4} lines at 5412\,\AA\ and 5471\,\AA, respectively. The line blend of \Ion{He}{2} 4686\,\AA\ and several \Ion{C}{4} lines scattered around 4660\,\AA\ is less useful, in the case of our LRS2-B data, because the splice of the UV and Orange channels located at 4645\,\AA\ corrupts the \Ion{C}{4} lines in the observations of some objects. For most objects, our best-fit models have He/C $\approx$ 1, but a few have He/C $\approx$ 3.7. The possible error in these ratios is about a factor of two. 

\begin{figure*}
\begin{center}
\includegraphics[width=0.95\linewidth]{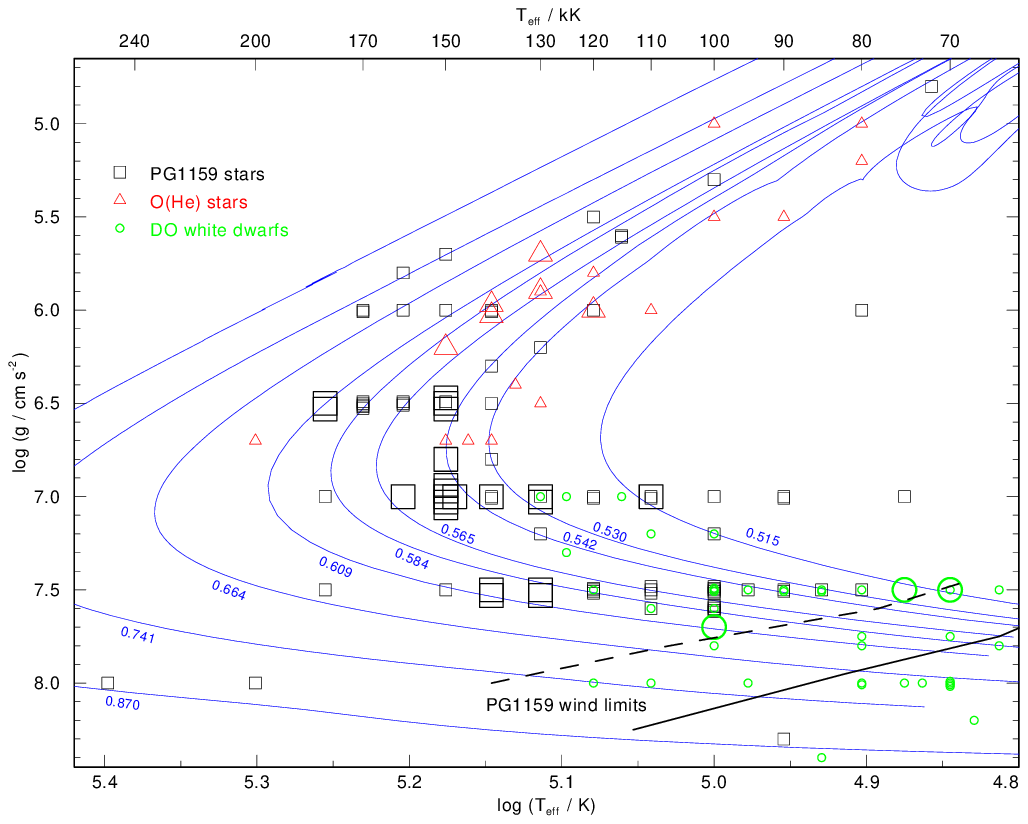}
\caption{Positions of our sample of H-deficient stars in the Kiel
  diagram (large symbols), together with all previously known objects (whether or not associated with a PN) of the spectral classes PG1159 and O(He), as well as a number of DO WDs (small symbols). 
  Evolutionary tracks (blue lines) for VLTP post-AGB stars labeled with the stellar mass in solar units are from \cite{MillerBertolamiAlthaus2006}. The solid and dashed black lines indicate two theoretical PG1159 wind limits, below which heavy elements settle out of the photosphere; for details see text.
}
\label{fig:gteff}
\end{center}
\end{figure*}

Concerning the oxygen abundance, we assumed O = 0.05 or 0.06 to calculate our models. This value is a rough mean of results of previous analyses. Only in two cases (Dr~37 and K 1-17) did we reduce the abundance to O = 0.02. The reason is that, at O = 0.05, the \Ion{O}{6} emission line at 5291\,\AA\ is accompanied by weak absorption wings, which are not observed.  Therefore, in general, we expect that the error of the O abundances given in Table~\ref{tab:parameters} is between 0.2 and 0.3~dex, depending on the quality of the spectrum. In the spectrum of Ek~3 we see no lines of oxygen {and an upper limit of O $\leq$ 0.06 is estimated.} The nitrogen abundance for Pa~28 is quite uncertain (by about 0.5~dex), because the emission strength of the \Ion{N}{5} 4945\,\AA\ line is rather sensitive to effective temperature and gravity. 

\subsection{O(He) stars and DOZ white dwarfs}

The spectra of the six O(He) stars and three DOZ WDs in our sample are dominated by \Ion{He}{2} absorption lines. The O(He) stars additionally exhibit an emission feature of \Ion{N}{5} at 4945\,\AA\ (less clear in StDr~162), similar, for example, to three recently discovered and analyzed field O(He) stars \citep{Jeffery2023}. This feature suggests that they have temperatures of around 140\,000\,K\null. We also identify the \Ion{N}{5} doublet in emission at 4604/4620\,\AA\ and another weak \Ion{N}{5} emission at 4520\,\AA\ (e.g., Kn~121). In contrast, we do not see lines from oxygen and silicon that were occasionally identified in other O(He) stars \citep[e.g., \Ion{O}{5}/\Ion{O}{6} and \Ion{Si}{4};][]{Reindl2014,werner14}. 
The spectrum of Pa~146 is contaminated by residual PN emission lines.

The spectra of our DO stars Pa~180 and StDr~61 exhibit weak \Ion{C}{4} absorption lines at 4660\,\AA\null. They are therefore classified as DOZ WDs, indicating trace amounts of carbon, as is often found in hot DOs that have a similar temperature and gravity \citep{werner14}. The DO star StDr~29 exhibits the \Ion{N}{5} doublet in absorption at 4604/4620\,\AA; hence, it is also classified as a DOZ\null. The presence of the \ion{C}{iv} lines at 4648 and 4660\,\AA\ in this star is uncertain, and we only determine an upper C limit.

\section{Evolutionary status}
\label{sect:discussion}

The positions of our PNNi in the Kiel diagram are displayed in Fig.\,\ref{fig:gteff}, where they are represented by the large plotting symbols. The small symbols mark previously known objects belonging to the same spectral classes. One obvious result is that our objects strongly tend to lie in the bottom half of this diagram, i.e., they are highly evolved, consistent with them belonging to faint PNe that are well into the process of dissipating into the ISM. 

Effective temperatures and gravities of the analyzed PG1159 stars range between \Teff = 110\,000--180\,000\,K and \logg = 6.5--7.5, placing them near the maximum temperatures of respective post-AGB evolutionary tracks. It is thought that further contraction and cooling transforms the PG1159 stars into DO WDs near the PG1159 wind limit.  
This limit, according to the theory of \citet{UnglaubbBues2000}, is indicated by the solid black line in Fig.\,\ref{fig:gteff}. The mass-loss rate of the radiation-driven wind at this position of the evolutionary tracks becomes so weak that gravitational settling becomes able to remove heavy elements from the WD atmosphere. Thus, no PG1159 stars are expected to be found at significantly cooler temperatures. The dashed line in Fig.\,\ref{fig:gteff} is the wind limit assuming a mass-loss rate that is ten times lower. Fittingly, all our stars lie above the dashed line.

Only one of our PG1159 stars, Pa~28,  exhibits \Ion{N}{5} (emission) lines, from which an abundance of about N = 0.01 can be inferred. The presence of nitrogen is a consequence of a VLTP, for which evolutionary models predict a complete burning of hydrogen and an enhancement of nitrogen, for a solar-metallicity star, of up to a few percent. In contrast, an LTP causes dilution of hydrogen, and the N abundance will be at most roughly 0.001 \citep[see, e.g.,][]{werner2006}. This means that all but one of our PG1159-type PNNi have experienced an LTP, i.e., a thermal pulse in the previous pre-WD phase. This is in line with the fact that they still have an observable PN, because a VLTP can occur in WDs with ages so high that their PN has dispersed long ago. 

All of our PG1159-type PNNi are located in the GW~Vir pulsational instability strip, close to its blue edge \citep{Sowicka2023}. Although this strip is not pure, in the sense that not all objects located within it actually exhibit pulsations, the chances are excellent that many of our stars might be found to be short-period $g$-mode pulsators (GW~Vir variables). Just 25 such pulsators (nine of them within a PN) are known; therefore, our new PG1159 stars could significantly increase these numbers. GW~Vir variables allow one to study the interior structure of the stars and their evolutionary rate \citep[e.g.,][]{Kepler2022}, and enable an independent mass determination \citep[e.g.,][]{Calcaferro2024}. In particular, they can be used to advance the mixing-length theory, a major ingredient in stellar-evolution modeling \citep{Ocampo2026}.

The temperatures and gravities of our O(He) stars are very similar (\Teff = 120\,000--150\,000\,K, \logg = 5.7--6.2) and their evolutionary state is before they reach their maximum effective temperature. They all have nitrogen abundances of N = 0.01. The three previously known O(He)-type PNNi (K~1-27, LoTr~4, Pa~5) have similar spectra (including the \Ion{N}{5} emissions and without C and O lines) and parameters, too \citep[\Teff = 120\,000--145\,000\,K, \logg = 5.8--6.7,][]{Reindl2014,DeMarco2015}. It is believed that the DO WDs observed before the wind limit in the Kiel diagram (Fig.\,\ref{fig:gteff}) are the immediate progeny of O(He) stars. A binary He-WD merger invoked for the formation scenario of O(He) stars cannot explain the existence of a PN around these stars. Even if a PN was ejected during the merger, it would have dispersed long ago. An alternative scenario would be a common-envelope (CE) ejection during a He-WD merger with the core of an AGB star \citep{Soker2013} because the post-merger timescales can be expected to be much shorter. Another possibility is the CE ejection by an in-spiral of a planet or a brown dwarf \citep{Soker1998} onto an AGB or red-giant branch star \citep[see discussion in][]{Reindl2014}. 

Our discovery of three DO WDs means that they could be the first identified PNNi of this spectral type. StDr~29 is ``likely'' a PN, according to the HASH database, and a comment there says that a nebular spectrum confirmed its PN nature. The star's effective temperature of \Teff = 100\,000\,K and gravity of \logg = 7.7 place the star on a cooling track with a remnant mass of 0.61\,\Msol\ and close to the PG1159 wind limit (Fig.\,\ref{fig:gteff}). It is probably the descendant of an O(He) star because, as has been pointed out by \cite{MMM2024}, the gravitational settling timescales for the transformation of a PG1159 star into a DO WD are much longer than the lifetime of a PN\null. StDr~61 and Pa~180 are classified as a ``possible PN'' and a ``new PN candidate,'' respectively, in the HASH database.

Using the theoretical evolutionary tracks of \cite{MillerBertolamiAlthaus2006}, as displayed in Fig.\,\ref{fig:gteff}, we derived the stellar masses listed in the final column of Table~\ref{tab:parameters}. They are in a narrow range of 0.51--0.62\,\Msol, with a mean of 0.56\,\Msol. The same mean mass value was determined for our sample of hydrogen-rich PNNi in Paper~VI, and is also in very good agreement with hydrogen-rich DAO WDs investigated by \citet[0.58\,\Msol]{2010ApJ...720..581G} and by \citet[0.55\,\Msol]{Filiz2024}. 

\section{Summary}
\label{sect:summary}

In conclusion, our spectroscopic survey has revealed 30 new H-deficient PNNi, for which we have performed the first classifications and atmospheric analyses. Our study reveals that all of them are extremely hot (\Teff = 70\,000--180\,000\,K) and have high surface gravities (\logg = 5.9--7.7). Our  survey significantly increases the number of PN nuclei belonging to the relatively rare hydrogen-deficient class. Out of our 30 newly classified objects, we found 21 PG1159 stars. Up until now, 71 PG1159 stars were known, of which 25 lie within a PN\null. This paper thus increases the total number of identified PG1159 stars by 30\%, and the number of PG1159-type PNNi by 84\%. 

We found six new O(He)-type PNNi. Up until now, a total of 14 O(He) stars was known, out of which only three are PNNi. We thus increased the total number of known O(He) stars by 43\%, and tripled the number of known O(He) PNNi. Additionally, we identified three DOZ WDs, the first objects of this spectral class found to be associated with a PN.

It is noteworthy that we did not find any [WR]-type PNNi, which are supposed to be (more luminous) progenitors of PG1159 stars \citep[but note that their locations in the Hertzsprung-Russell diagram overlap; see, e.g., the discussion in][]{WernerBondIII2024}. The probable reason is that our spectroscopic survey is primarily aimed at relatively faint and extended PNe, which tend to harbor evolved nuclei of low luminosity. 

Finally, our spectroscopic survey has also revealed a large number of new hydrogen-rich PNNi, the more common spectral group. Their identification, classification, and analysis will be the subject of a forthcoming publication in this series.

\begin{acknowledgements}

As this paper was being completed, we learned that our colleague Detlef Sch\"onberner passed away on 2026 February~4. His pioneering specctroscopic investigations of the nuclei of faint planetary nebulae, e.g., \citet{Napiwotzki1995}, were a main source of inspiration for our project.

This paper is based on observations obtained with the Hobby-Eberly Telescope (HET), which is a joint project of the University of Texas at Austin, the Pennsylvania State University, Ludwig-Maximilians-Universit\"at M\"unchen, and Georg-August Universit\"at G\"ottingen. The HET is named in honor of its principal benefactors, William P. Hobby and Robert E. Eberly. We thank the HET queue schedulers and nighttime observers at McDonald Observatory for obtaining the data discussed here.
The Low-Resolution Spectrograph 2 (LRS2) was developed and funded by The University of Texas at Austin McDonald Observatory and Department of Astronomy, and by The Pennsylvania State University. We thank the Leibniz-Institut f\"ur Astrophysik Potsdam (AIP) and the Institut f\"ur Astrophysik G\"ottingen (IAG) for their contributions to the construction of the integral-field units.

We acknowledge the Texas Advanced Computing Center (TACC) at The University of Texas at Austin for providing high-performance computing, visualization, and storage resources that have contributed to the results reported within this paper.

This work has made use of data from the European Space Agency (ESA) mission
{\it Gaia\/} (\url{https://www.cosmos.esa.int/gaia}), processed by the {\it Gaia\/}
Data Processing and Analysis Consortium (DPAC,
\url{https://www.cosmos.esa.int/web/gaia/dpac/consortium}). Funding for the DPAC
has been provided by national institutions, in particular the institutions
participating in the {\it Gaia\/} Multilateral Agreement.

This research has made use of the SIMBAD database, operated at CDS, Strasbourg, France.

Based on observations made with the NASA {\it Galaxy Evolution Explorer}.
\GALEX\/ was operated for NASA by the California Institute of Technology under NASA contract NAS5-98034.

The Pan-STARRS1 Surveys (PS1) and the PS1 public science archive have been made possible through contributions by the Institute for Astronomy, the University of Hawaii, the Pan-STARRS Project Office, the Max-Planck Society and its participating institutes, the Max Planck Institute for Astronomy, Heidelberg and the Max Planck Institute for Extraterrestrial Physics, Garching, The Johns Hopkins University, Durham University, the University of Edinburgh, the Queen's University Belfast, the Harvard-Smithsonian Center for Astrophysics, the Las Cumbres Observatory Global Telescope Network Incorporated, the National Central University of Taiwan, the Space Telescope Science Institute, the National Aeronautics and Space Administration under Grant No.\ NNX08AR22G issued through the Planetary Science Division of the NASA Science Mission Directorate, the National Science Foundation Grant No. AST-1238877, the University of Maryland, Eotvos Lorand University (ELTE), the Los Alamos National Laboratory, and the Gordon and Betty Moore Foundation.

We thank several amateur colleagues, including Peter Goodhew, Dana Patchick, Sakib Rasool, Jon Talbot, and others, for pointing out interesting discoveries of new, faint PNe; and we congratulate them on their amazing deep imaging.

\end{acknowledgements}

\bibliographystyle{aa}
\bibliography{aa}


\begin{appendix}


\section{Additional table and figures}

\FloatBarrier

\begin{table*}[] 
\small
\centering
\caption{Log of HET LRS2-B spectroscopic observations.}
\label{tab:observations}
\begin{tabular}{lccclcc}
\hline 
\hline 
\noalign{\smallskip}
Nucleus of & Date & Exposure & & Nucleus of & Date & Exposure \\
           & [YYYY-MM-DD] & [s] & & & [YYYY-MM-DD] & [s] \\
\hline
\noalign{\smallskip}
Abell 52      & 2024-08-17 & $2\times506$    & & NGC 6765        & 2024-10-18 & $907   $ \\  
              & 2024-09-18 & $2\times508$       & &              & 2024-11-01 & $1007  $ \\  
              & 2024-09-20 & $2\times508$       & & NHZ 2        & 2025-07-27 & $1207  $ \\  
              & 2024-11-03 & $2\times506$       & &              & 2025-09-09 & $1206  $ \\  
              & 2025-05-09 & $2\times507$       & &              & 2025-10-26 & $1207  $ \\  
Alves 5       & 2025-02-17 & $407$      & & Ou 7         & 2024-10-06 & $2\times808   $ \\  
              & 2025-03-21 & $407$      & &              & 2024-11-17 & $2\times807   $ \\  
              & 2025-03-31 & $406$      & &              & 2024-11-21 & $2\times808   $ \\  
              & 2025-04-07 & $608$      & &              & 2025-08-25 & $2\times808   $ \\  
Dr 37         & 2025-06-28 & $2\times1207  $ & & Pa 28   & 2024-10-11 & $2\times1507  $ \\      
              & 2025-10-22 & $2\times1358  $ & &                 & 2024-11-17 & $2\times1507  $ \\         
Ek 3          & 2025-10-11 & $2\times1008  $ & & Pa 144  & 2025-07-28 & $2\times1007  $ \\           
Fal 3         & 2024-09-13 & $407     $ & & Pa 146       & 2024-10-03 & $608     $ \\         
              & 2024-09-20 & $408     $ & &              & 2025-04-13 & $606     $ \\    
              & 2025-03-30 & $407     $ & &              & 2025-08-03 & $607     $ \\    
              & 2025-03-31 & $407     $ & & Pa 153       & 2024-08-25 & $907     $ \\     
              & 2025-04-13 & $407     $ & &              & 2024-09-22 & $2\times508   $ \\          
Fal Objet 1   & 2024-10-25 & $246     $ & &              & 2024-11-15 & $2\times508   $ \\          
              & 2024-12-09 & $247     $ & &              & 2024-11-30 & $2\times507   $ \\             
              & 2025-01-12 & $248     $ & &              & 2024-12-18 & $2\times507   $ \\            
              & 2025-01-14 & $247     $ & &              & 2025-09-11 & $2\times507   $ \\              
HaWe 15       & 2024-08-09 & $2\times608   $ & &                 & 2026-01-04 & $1008  $ \\                  
              & 2024-09-10 & $2\times607   $ & &                 & 2026-01-20 & $1006  $ \\                           
              & 2024-11-04 & $2\times607   $ & & Pa 180  & 2025-07-08 & $907     $ \\        
              & 2025-08-02 & $2\times607   $ & &                 & 2025-08-13 & $907   $ \\          
K 1-17        & 2024-09-11 & $2\times606   $ & & Pa J0637+3327 & 2024-09-20 & $607   $ \\        
              & 2024-09-28 & $2\times608   $ & &                 & 2024-09-28 & $607   $ \\         
              & 2024-11-04 & $606     $ & &              & 2024-11-07 & $608     $ \\     
              & 2024-11-07 & $2\times609   $ & &                 & 2024-11-21 & $907   $ \\           
              & 2025-04-09 & $757     $ & &              & 2024-12-12 & $907     $ \\         
              & 2025-04-09 & $69      $ & &              & 2025-01-01 & $907     $ \\        
              & 2025-06-08 & $2\times758   $ & & PFP 1   & 2024-10-24 & $248     $ \\          
              & 2025-06-16 & $2\times907   $ & &                 & 2025-01-02 & $247   $ \\       
              & 2025-07-28 & $2\times760   $ & &                 & 2025-01-12 & $247   $ \\         
              & 2025-07-29 & $2\times756   $ & &                 & 2025-10-29 & $308   $ \\        
K 2-1         & 2024-10-14 & $2\times607   $ & & StDr 13         & 2022-12-28 & $2\times758   $ \\     
              & 2024-10-24 & $2\times606   $ & &                 & 2024-11-22 & $2\times1008  $ \\         
              & 2024-11-15 & $2\times607   $ & &                 & 2024-12-08 & $2\times1007  $ \\    
Kn 58         & 2024-09-29 & $2\times1507  $ & & StDr 29         & 2024-08-14 & $908   $ \\           
              & 2024-10-12 & $2\times1507  $ & &                 & 2025-08-14 & $906   $ \\              
              & 2024-11-20 & $2\times1507  $ & &                 & 2026-02-22 & $907   $ \\             
              & 2024-11-30 & $2\times1507  $ & & StDr 61         & 2025-08-07 & $187   $ \\              
              & 2024-12-25 & $2\times1507  $ & &                 & 2025-12-14 & $307   $ \\             
Kn 62         & 2024-09-26 & $2\times1006  $ & & StDr 144        & 2025-08-29 & $456   $ \\              
              & 2025-01-04 & $2\times1007  $ & &                 & 2025-12-24 & $457   $ \\            
              & 2025-01-20 & $2\times1008  $ & &                 & 2026-01-20 & $457   $ \\          
              & 2025-09-24 & $2\times1007  $ & & StDr 162        & 2025-09-26 & $2\times1358  $ \\        
Kn 63         & 2024-09-29 & $907     $ & & TaWe 1       & 2024-10-30 & $2\times1007  $ \\          
              & 2025-01-01 & $908     $ & &              & 2024-11-27 & $2\times1007  $ \\           
              & 2025-01-11 & $908     $ & &              & 2025-10-20 & $2\times1007  $ \\         
              & 2025-10-19 & $1207    $ & & WHTZ 1       & 2024-08-22 & $367     $ \\           
Kn 121        & 2024-08-11 & $247     $ & &              & 2024-09-12 & $507     $ \\         
              & 2024-08-30 & $247     $ & &              & 2024-09-18 & $368     $ \\          
              & 2025-04-21 & $247     $ & &              & 2025-04-11 & $507     $ \\            
\noalign{\smallskip}
\hline
\end{tabular} 
\end{table*}


\begin{landscape}
\addtolength{\textwidth}{6.3cm} 
\addtolength{\evensidemargin}{0cm}
\addtolength{\oddsidemargin}{0cm}
\begin{figure*}[]
  \centering  
  \includegraphics[width=0.68\textwidth,angle=-90]{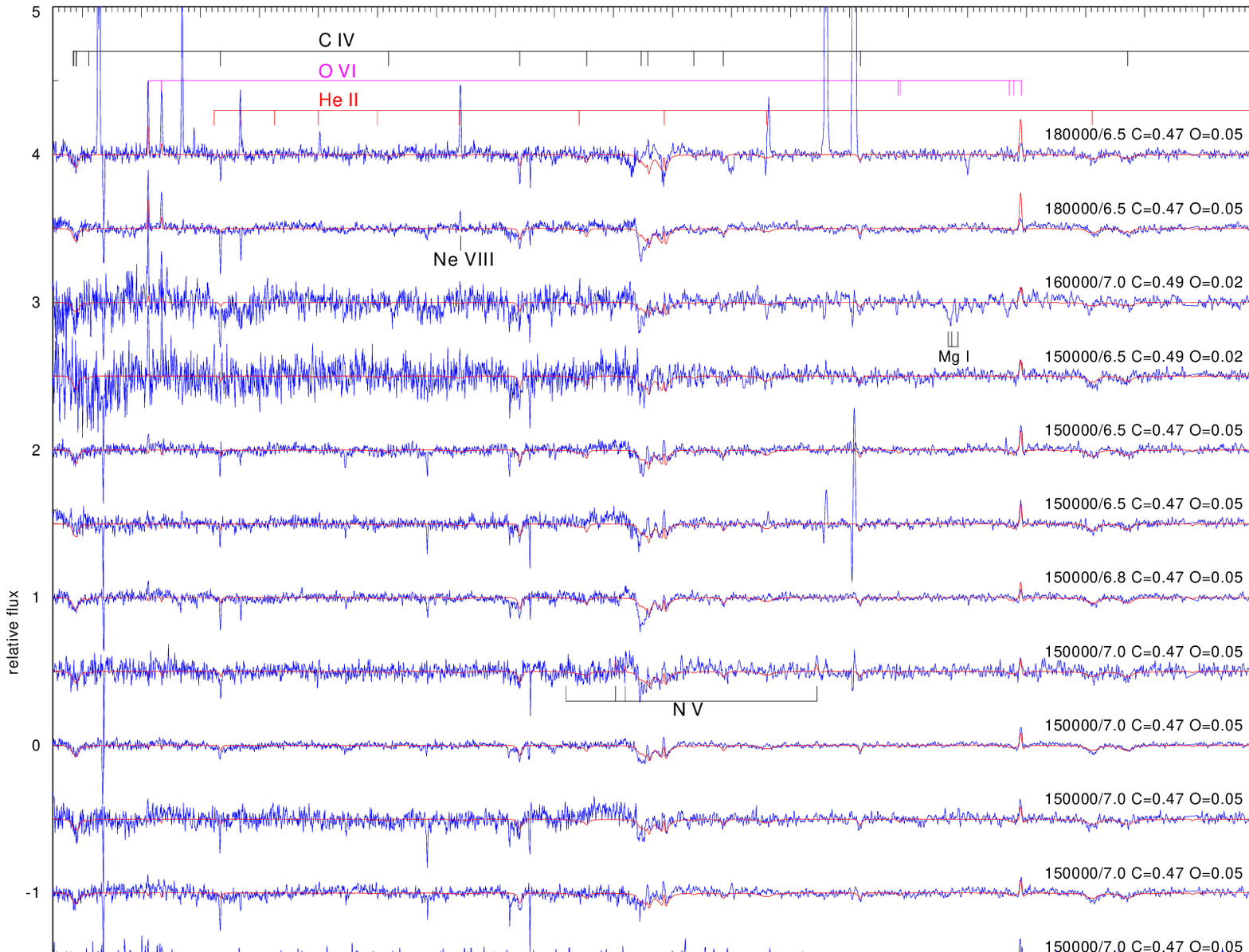}
  \caption{HET spectra of 15 of our PG1159-type central stars (blue graphs), ordered by decreasing \Teff. At each spectrum we give the PN name and the spectral type. Overplotted in red are the best-fit models, whose parameters are indicated (\Teff, \logg, metal abundances in mass fractions). Photospheric, interstellar, PN, telluric, and sky lines are marked, as well as detector artefacts. Distortions in the continuum levels around 4645~\AA\ are due to the splice between the UV and Orange spectrograph channels (see text).}
\label{fig:pg1159}
\end{figure*}
\end{landscape}

\begin{landscape}
\addtolength{\textwidth}{6.3cm} 
\addtolength{\evensidemargin}{0cm}
\addtolength{\oddsidemargin}{0cm}
\begin{figure*}[]
  \centering  
  \includegraphics[width=0.68\textwidth,angle=-90]{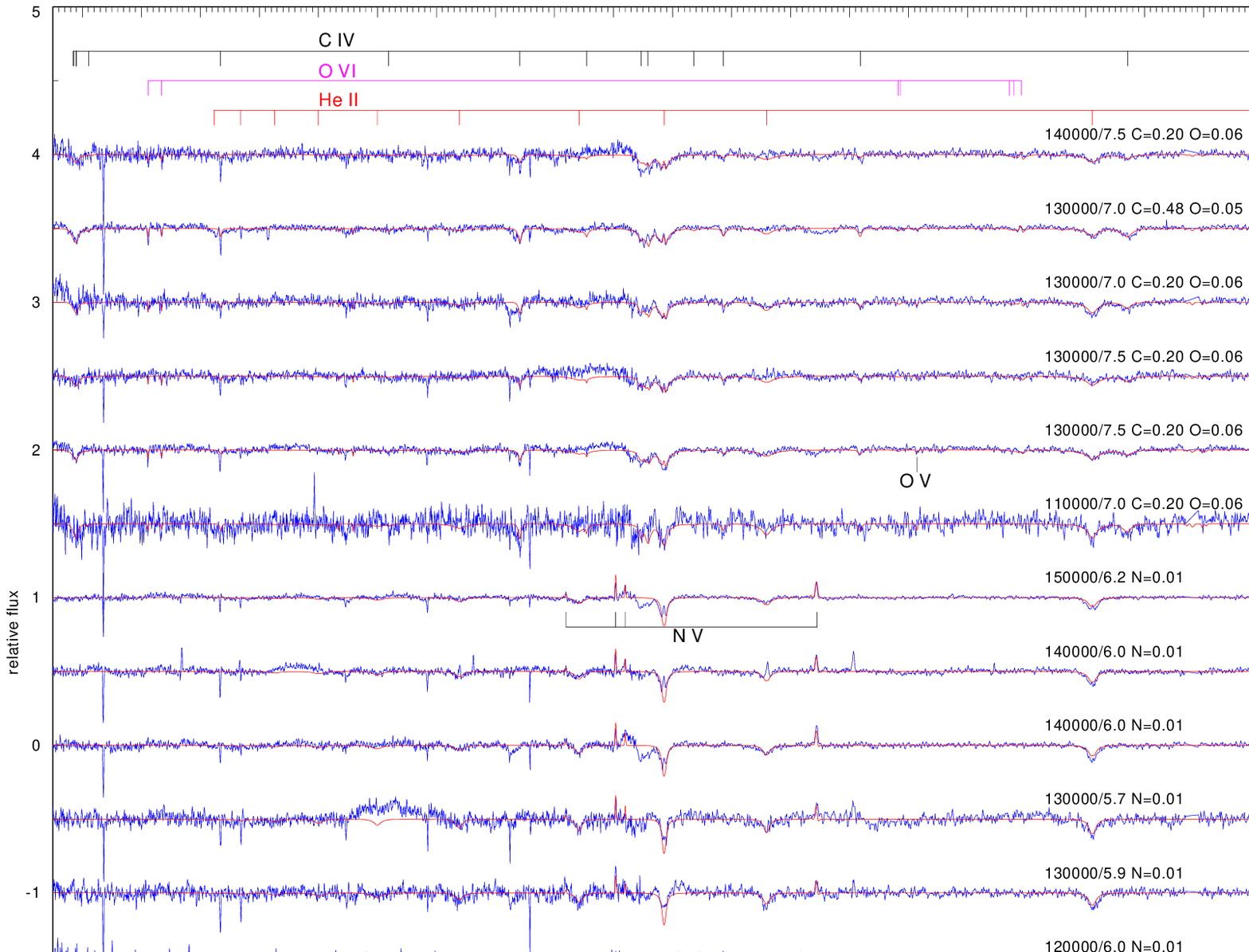}
  \caption{Same as Fig.\,\ref{fig:pg1159} but for the remaining six PG1159 stars, plus the six O(He) stars and the three DOZ WDs.}
\label{fig:ohe}
\end{figure*}
\end{landscape}

\end{appendix}

\end{document}